\documentclass[reprint,secnumarabic,amssymb, nobibnotes, aps, prx]{revtex4-1}

\usepackage{color}
\usepackage{graphics, graphicx, subfigure, wrapfig}
\usepackage{graphicx}
\usepackage{verbatim, multirow}
\usepackage{dcolumn}
\usepackage{bm}
\usepackage{amsmath, braket}
\usepackage{array}
\usepackage{xr}
\newcolumntype{X}[1]{>{\centering\arraybackslash\hspace{0pt}}p{#1}}
\newcolumntype{M}[1]{ >{\centering\arraybackslash}m{#1}}
\newcommand{\roml}[1]{\lowercase\expandafter{\romannumeral #1\relax}}
\newcommand{\romu}[1]{\uppercase\expandafter{\romannumeral #1\relax}}

\begin{document}

\title{Atomic positions independent descriptor for machine learning of material properties}

\author{Ankit Jain}
\email{ankitj@alumni.cmu.edu}
\affiliation{Department of Chemical Engineering, Stanford University, Stanford, California 94305, United States}
\author{Thomas Bligaard}
\email{bligaard@slac.stanford.edu}
\affiliation{SUNCAT Center for Interface Science and Catalysis, SLAC, National Accelerator Laboratory, 2575 Sand Hill Road, Menlo Park, California, 94025, United States}

\date{\today}%


\begin{abstract}
The high-throughput screening of periodic inorganic solids using machine learning
methods requires atomic positions to encode structural
and compositional details into appropriate material descriptors. These atomic positions are not available {\it a priori} for new materials which severely limits exploration of novel materials. We overcome this limitation by using only crystallographic symmetry information in the structural description of materials. We show that for materials with identical structural symmetry, machine learning is trivial and accuracies similar to that of density functional theory calculations can be achieved by using only atomic numbers in the material description. For machine learning of formation energies of bulk crystalline solids, this simple material descriptor is able to achieve prediction mean absolute errors of only $0.07$ eV/atom on a test dataset consisting of more than 85,000 diverse materials. This atomic-position independent material descriptor presents a new route of materials discovery wherein millions of materials can be screened by training a machine learning model over a drastically reduced subspace of materials.
\end{abstract}

\maketitle

{\bf Introduction:}
Machine learning (ML) methods are now able to predict material properties with similar accuracy as those of density functional theory (DFT) calculations but with several orders of magnitude reduced computational cost\cite{hong2013, fujimura2013, deml2014, carrete2014, seko2014, kirklin2015}. ML methods require computational representation of materials which is achieved by mapping structure and composition to descriptors\cite{faber2015, ghiringhelli2015}. The majority of descriptors used in ML of inorganic periodic solids, such as the extended Coulomb matrix\cite{faber2015} and the partial radial distribution function\cite{schutt2014}, require atomic positions. This explicit use of atomic coordinates depends on prior knowledge of accurate positions for all atoms in a material with limited applicability to new and/or unknown materials (where atomic positions are not known {\it a priori}, as in the case of high-throughput calculations). Recently, there have been attempts to reduce this dependence of structure description on atomic coordinates, for instance, with a lattice volume independent Voronoi-tessellation (VT) based descriptor\cite{ward2017} and perturbation-tolerant crystal graph (CG) based convolution neural network (CNN)\cite{xie2018}, but a true atomic positions independent (apI) descriptor capable of describing structure details without explicit use of atomic positions is still non-existent.

Here, we present a crystallographic method for apI-description of structural details for inorganic periodic solids. Using this method, materials are clustered into different structure types on the basis of symmetry. Within a given structure-type cluster only material composition is  needed to accurately describe the system.  We first show that we can use an apI descriptor effectively within a structure type, and we then develop a universal apI descriptor (U-apI) that extends across structure types.

For the prediction of material's formation energies (FE) relative to their standard elemental references, accuracies similar to that of DFT calculations are achieved within these structure-type clusters by using a representation learning feedforward neural network (RLFNN) employing only atomic numbers of participating species. For FE predictions on the 20 most frequently appearing structure types from the open quantum materials database (OQMD)\cite{kirklin2015}, an attention-based convolution neural network (ABCNN) employing U-apI descriptor is able to match apD descriptor based ML models by delivering a regularized performance with an average test mean absolute error (MAE) of  $0.07$ eV/atom.

The U-apI descriptor based ABCNN is the first-ever apI descriptor based material ML model which is capable of learning across different structure types while matching the performance of more involved apD descriptor based ML models. The U-apI descriptor based ABCNN presents a new paradigm of high-throughput material discovery where millions of compounds can be reliably screened using ML without relying on atomic coordinates.

{\bf Structure Types:}
Three-dimensional periodic solids can be classified into one of the 14 Bravais lattices\cite{hahn1983}. These 14 Bravais lattices, when combined with 32 point groups,
result in 230 three-dimensional space groups (without considering spin)\cite{fedorov1971, schoenflies1923}. Each of these 230 space groups is divided into conjugate subgroups called Wyckoff sites\cite{wyckoff1922}. Atoms in crystals can sit only on special positions, called Wyckoff positions, which are points belonging to the Wyckoff site of the space group of the crystal\cite{downs2003}. For some Wyckoff sites all internal fractional coordinates are fixed, at the others one or more fractional coordinates can be changed while maintaining the symmetry of the Wyckoff site.

The atomic structure of a periodic solid is uniquely identified by specifying (\roml{1}) the crystal space group, (\roml{2}) chemical species, (\roml{3}) Wyckoff sites occupied by atoms, (\roml{4}) the unit-cell parameters, and (\roml{5}) numerical values of the fractional coordinates for occupied Wyckoff sites (needed for sites with free fractional coordinates). Fixing the space group and Wyckoff site of a material uniquely set the symmetry environment of the atoms. Herein, this will be referred to as structure type. Specifying the chemical species that occupy specific Wyckoff sites results in a completely specified structure. Within this fixed chemical and symmetry environment, the numerical values of the unit-cell parameters and free fractional coordinates can be determined by performing DFT calculations or by experimental synthesis and characterization of material.

\begin{figure}
\begin{center}
\includegraphics{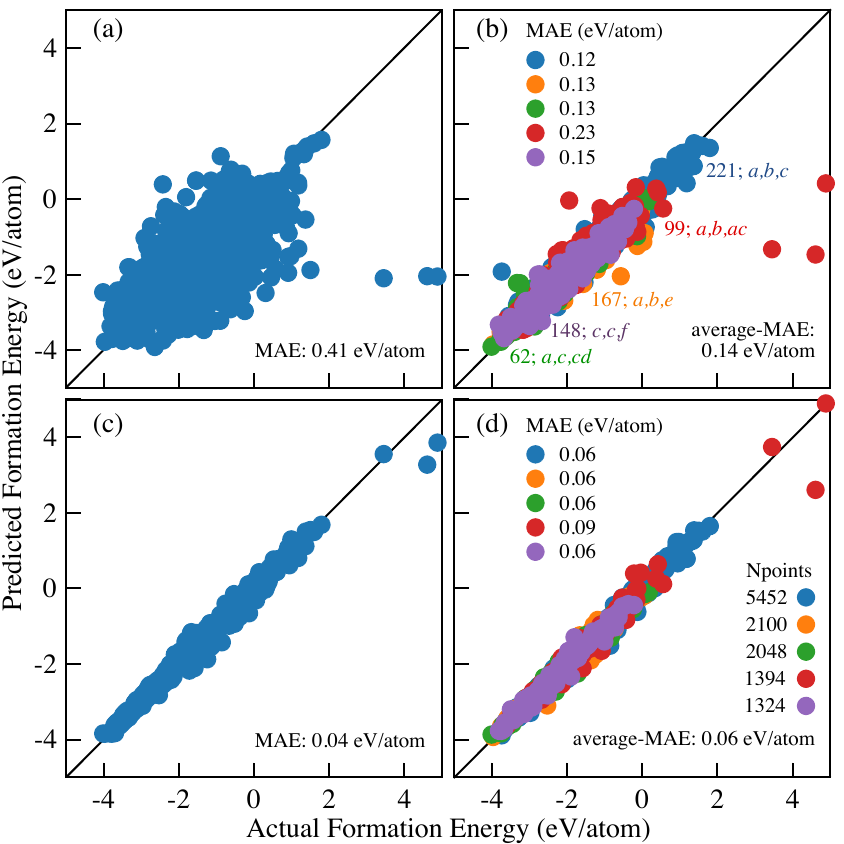}
\end{center}
\caption{{\bf Role of structural information in the ML of materials.}
The predicted FEs of 12,318 $\text{ABC}_3$ stoichiometry materials from OQMD. In (a) and (b), materials are described using a simple composition-only descriptor
consisting of 15 elemental properties of species A, B, and C by training the ML model over all structure types simultaneously in (a) and separately in (b).
In (c) and (d), predictions are made using VT-based apD-descriptor\cite{ward2017} by training the ML model over all structure types simultaneously in (c) and separately in (d). The labels next to data points in (b) show the structure-type
(space group; Wyckoff-sites). All predictions are made using a random forest ML algorithm with 10 decision trees. Only test data is shown (20\% of data).}
\label{fig_validation}
\end{figure}

{\bf ML within Structure Type:}
We begin our apI-descriptor design with a simple composition-only descriptor consisting of 15 elemental properties of participating species\cite{atomic_properties}.  The ML is performed using the random forest (RF) ML algorithm with 10 decision trees (as implemented in the ML library SKlearn\cite{sklearn}). The ML is performed on 12,318 $\text{ABC}_3$ stoichiometry materials from the OQMD database for the prediction of FEs of materials for decomposition into elemental standard states \cite{ward2017}.
The materials are randomly assigned into training and test datasets consisting of 80\% and 20\% of the total materials, respectively. 
Fig.~\ref{fig_validation} reports the predicted FEs of the materials in the test dataset materials. A MAE of 0.41 eV/atom is achieved by using a simple composition-only descriptor, see Fig.~\ref{fig_validation}(a).
The large MAE is due to the presence of five different $\text{ABC}_3$ allotropes in the dataset;  indistinguishable by elemental properties alone. To distinguish these allotropes, materials are pre-clustered on the basis of structure type before performing ML [Fig.~\ref{fig_validation}(b)] reducing the average MAE to $0.14$ eV/atom. This three-fold reduction in the MAE with structure-type clustering demonstrates
the importance of retaining structural details. This is in contrast with \cite{legrain2017} where authors suggested similar ML performance from structure-dependent and structure-independent descriptors for vibrational free energy and entropy predictions.


In Fig.~\ref{fig_validation}(c), the VT-based apD descriptor\cite{ward2017} is used for the material description and the prediction MAE is $0.04$ eV/atom. In comparison to structure-type pre-clustering based apI descriptor in Fig.~\ref{fig_validation}(b), the superior performance of VT-based apD descriptor in Fig.~\ref{fig_validation}(c) can originate from: (\roml{1}) simultaneous training over all structure-types, (\roml{2}) additional apD details about fractional volumes of atoms, and (\roml{3}) inclusion of 271 elemental/structural properties of species [opposed to 15 per species in Fig.~\ref{fig_validation}(b)]. The effects of (\roml{1})-(\roml{3}) are investigated independently in following paragraphs.

The effect of (\roml{1}) is tested by separately training the VT-based apD descriptor ML model on the different structure types  [Fig.~\ref{fig_validation}(d)]. The average MAE increases only marginally to $0.06$ eV/atom. The minimum MAE across different structure types is $0.06$ eV/atom in Fig.~\ref{fig_validation}(d) compared to an average MAE of $0.04$ eV/atom in Fig.~\ref{fig_validation}(c) showing that all structure types gained accuracy improvements by simultaneous training of the ML model in Fig.~\ref{fig_validation}(c).

The effect of including additional structural details [(\roml{2} from above)] was studied by focusing on the perovskite structure-type with space group 221 and Wyckoff sites $a$, $b$, and $c$ [blue data points in Figs.~\ref{fig_validation}(b) and \ref{fig_validation}(d)]. For the perovskite structure-type, the only free parameter is the lattice constant. Therefore, the use of fractional atomic volume in the VT-based apD descriptor does not provide additional information beyond that already encoded in the Wyckoff-sites. The prediction MAE with structure-type pre-clustering based apI descriptor of $0.12$ eV/atom in Fig.~\ref{fig_validation}(b) compared to only $0.06$ eV/atom in Fig.~\ref{fig_validation}(d) rules out the additional contribution from apD structure details as the source of the superior performance of VT-based apD descriptor.

\begin{figure}
\begin{center}
\includegraphics{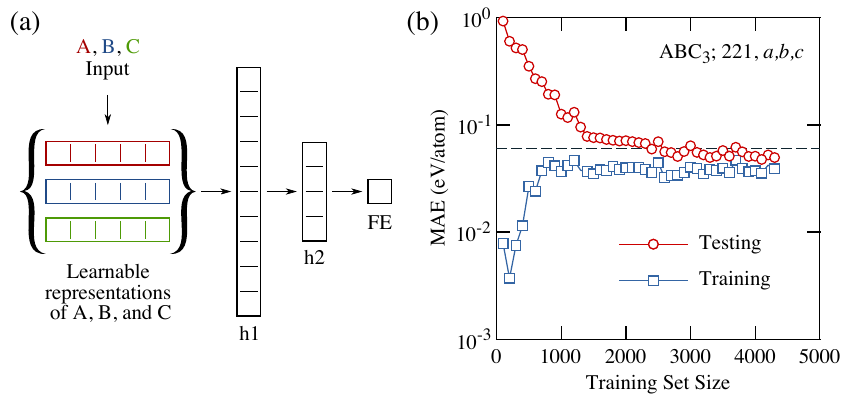}
\end{center}
\caption{{\bf ML for same structure-type materials.} (a) A representation-learning feedforward neural network employing atomic numbers of participating species for FE prediction of perovskite structure-type materials. (b) The effect of training set size on the prediction performance of the neural network. The network takes atomic numbers of species A, B, and C as an input and learns their five-dimensional representation from the training data.}
\label{fig_NN_split}
\end{figure}

To confirm that the inferior performance of the structure-type pre-clustering based apI descriptor in Fig.~\ref{fig_validation}(b) originates from non-optimized elemental properties of relevant species [factor (\roml{3}) from above], a RLFNN is employed for the ML of FEs of materials belonging to perovskite structure type in Fig.~\ref{fig_NN_split}. Instead of using pre-tabulated elemental properties, the RLFNN was designed to learn a five-dimensional representation of species from the training data itself. The RLFNN architecture is reported in Fig.~\ref{fig_NN_split}(a) where learnable  representations of species A, B, and C are passed through two fully connected hidden layers consisting of 10 and 4 neutrons to make the FE prediction. With this network design and for 82 chemical species in the perovskite structure-type from OQMD, the number of trainable weights are only 645 of which  410 weights are for elemental representations.

In Fig.~\ref{fig_NN_split}(b), the effect of training dataset size on predictability is presented by training the network on varying size datasets and evaluating the performance on $\sim$1100 test data points. For fewer than 1500 points in the training dataset, the network results in over-fitting indicated by large differences in the MAEs between training and testing datasets (regularization is not used in this network). With more than 2500 points in the training dataset, the network is able to achieve a prediction MAE of less than $0.06$ eV/atom, the same as that from VT apD descriptor based RF model in Fig.~\ref{fig_validation}(d).

\begin{figure*}
\begin{center}
\includegraphics{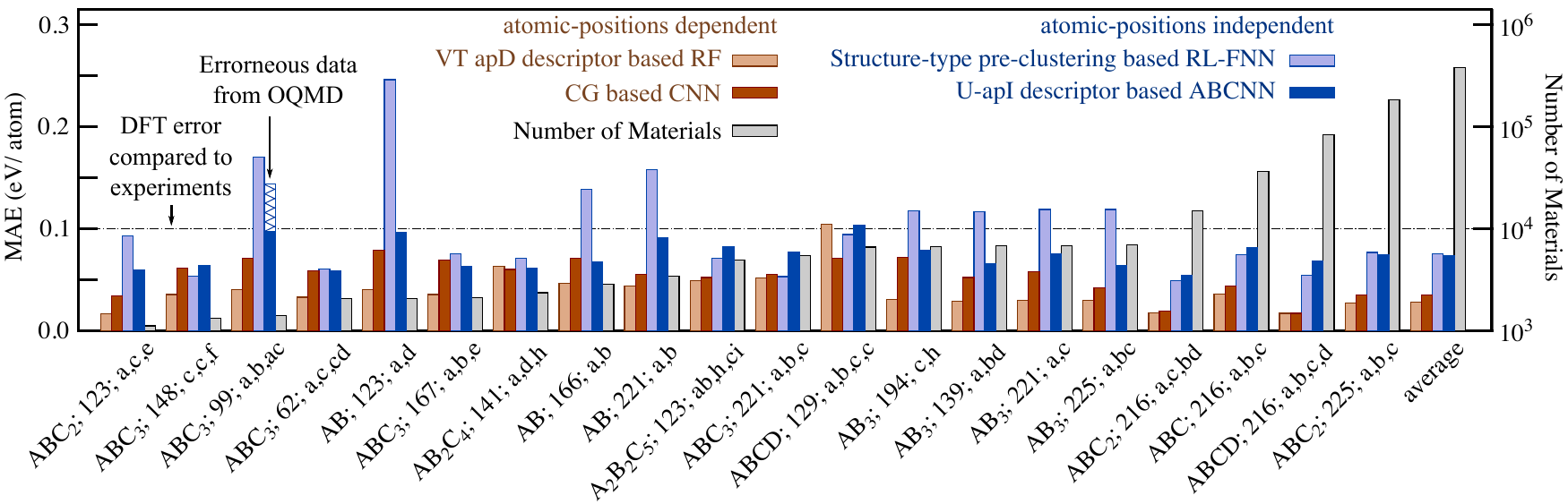}
\end{center}
\caption{{\bf ML across structure types.}
Comparison of atomic-positions dependent and atomic-positions independent descriptors based ML models for the prediction of FE of 20 most frequently appearing structure-types from OQMD. For atomic-positions dependent descriptors, VT apD descriptor based RF\cite{ward2017} and CG based CNN\cite{xie2018} are considered. For atomic-positions independent descriptors, structure-type pre-clustering based RLFNN and U-apI descriptor based ABCNN are considered. The x-axis labels represent material's stoichiometry; space group number; Wyckoff sites. For structure-type pre-clustering based RLFNN, a different ML model is trained on each structure type.}
\label{fig_top_prototype}
\end{figure*}

Fig.~\ref{fig_top_prototype} reports the performance of the structure-type pre-clustering apI descriptor based RLFNN to the VT apD descriptor based RF model for the 20 most-frequently appearing structure types in OQMD in Fig.~\ref{fig_top_prototype}. These structure types consist of more than $380,000$ materials from 13 space groups involving 41 different Wyckoff-sites and 84 elements. For the RLFNN a different neural network [similar to that in Fig.~\ref{fig_NN_split}(a)] is trained on different structure type. For the VT apD descriptor based RF model a single RF was trained by randomly collecting 80\% of the data from each structure type.

For structure types consisting of more than 3500 materials, RLFNN results in MAEs of less than $0.12$ eV/atom which is similar to the MAE for DFT calculations validated against experiments\cite{kirklin2015}. For structure types with fewer materials the MAEs from RLFNN increase up to a maximum of $0.25$ eV/atom. In comparison, the VT apD descriptor based RF model delivers a more regularized performance with maximum MAEs of only $0.10$ eV/atom across all considered structure types.

The RLFNN of Fig.~\ref{fig_NN_split}(a) is designed to learn the elemental representations of species from the training dataset. As shown in Fig.~\ref{fig_NN_split}(b), depending on the structure type, this requires an excess of 2500-3500 training data points. For densely populated structure types, while this trivial composition-only descriptor simplifies the material description and learning, for sparsely populated structure types, the prediction performance can be improved by sharing elemental information across structure types. To achieve this, we next focus our attention on the U-apI descriptor which is capable of learning across different structure types.

\begin{figure}
\begin{center}
\includegraphics{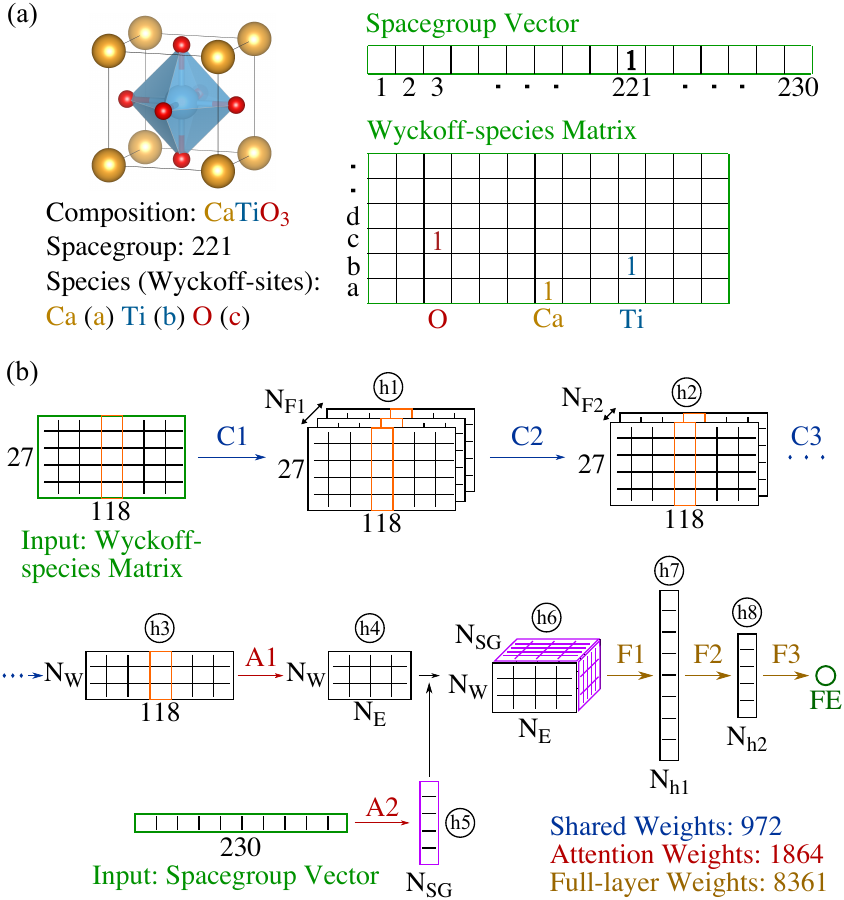}
\end{center}
\caption{
{\bf U-apI descriptor and attention-based convolution neural network.} (a) Illustration of U-apI descriptor (consisting of space group vector and Wyckoff-species matrix) for $\text{CaTiO}_3$ in the perovskite structure type. Only non-zero entries are explicitly listed for simplicity.
(b) An attention-based convolution neural network for the prediction of material properties using the U-apI descriptor. The shared, attention, and fully-connected layers are represented by blue, red, and brown colors. $\text{F}1$, $\text{F}2$, and $\text{F}3$ represents $(W1,b1)$, $(W2,b2)$, and $(W3,b3)$ respectively. The total number of trainable weights in the network are 11197 which are obtained by setting the values of $N_{F1}$, $N_{F2}$, $N_W$, $N_E$, $N_{SG}$, $N_{h1}$, and $N_{h2}$ at 4, 4, 4, 8, 4, 60, and 10 respectively.}
\label{fig_NN}
\end{figure}

{\bf ML across structure types:}
As presented in Figs.~\ref{fig_validation}(a) and \ref{fig_validation}(b),  space group and Wyckoff-sites are needed in the material description (along with the material composition) for learning across different structure-types. While both space group and Wyckoff-sites can be represented using simple scalars in the material description, this would suggest, for instance, that space groups 10 and 220 can be combined to form space group 230. To avoid this misrepresentation, space groups in U-apI descriptor are represented by a one-hot vector of length 230. The occupied Wyckoff-sites and corresponding atomic species are represented by a matrix of size $27\times118$ (referred to as Wyckoff-species matrix in this work), where 27 is the maximum number of Wyckoff sites in any space group\cite{aroyo2006} and 118 is the number of elements from the periodic table. The  element $(i,j)$ of the Wyckoff-species matrix denotes the number of Wyckoff sites of type $i$ occupied by an element of atomic number $j$. The U-apI descriptor for $\text{CaTiO}_3$ in the perovskite structure type is illustrated in Fig.~\ref{fig_NN}(a).

The simple concatenation of the space group vector and  Wyckoff-species matrix results in a tensor of size $230\times27\times118$. Training using simple ML models, such as linear regression and feedforward neural network, on this gigantic tensor, produces a minimum of 732,780 fitting parameters which are more than the number of entries in modern high-throughput material databases; thereby potentially causing over-fitting. To avoid this, U-apI descriptor is used with weights-sharing ABCNN\cite{yin2015} as presented in Fig.~\ref{fig_NN}(b). In this neural network, the columns of Wyckoff-species matrix are passed through three weights-sharing layers followed by a multiplication with species attention weights to get an embedded representation of Wyckoff-species matrix. This embedded representation is next weighted by space group attention weights before finally passing through two consecutive fully connected hidden layers to make a prediction for the material property.

Mathematically, the input Wyckoff-species matrix, $W_{i,j}$, is passed through two convolution layers consisting of $N_{F1}$ and $N_{F2}$ filters followed by a weights-sharing layer to obtain a $N_W$ length embedded representation of columns of Wyckoff-species matrix  as:
\begin{eqnarray}
h1_{i,j}^{k} &=& f(C1_{i}^{k} \times W_{i,j}), \label{eqn_conv1} \\
h2_{i,j}^{l} &=& f(\sum_k C2_{i}^{k,l} \times h1_{i,j}^{k}), \label{eqn_conv2} \\
h3_{m,j} &=& f(\sum_{i,l} C3_{i,l}^{m} \times h2_{i,j}^{l}), \label{eqn_conv3}
\end{eqnarray}
where $k$, $l$, and $m$ loop from 1 to $N_{F1}$, $N_{F2}$, and $N_W$, $f()$ is the activation function introducing non-linearity in the network, and $C1$, $C2$, and $C3$ are trainable convolution filter weights. The obtained embedded representation of Wyckoff-species matrix is next weighted by elements and space group attention weights to obtain the hidden material representation $h6_{m,n,q}$ as:
\begin{eqnarray}
h4_{m,n} &=& \sum_{j} A1_{j,n} \times h3_{m,j} \label{eqn_attn1}, \\
h5_{q} &=& \sum_{p} A2_{p,q} \times S_{p} \label{eqn_attn2}, \\
h6_{m,n,q} &=& h4_{m,n} \times h5_{q} \label{eqn_final_embdd},
\end{eqnarray}
where $S_p$ is the input space group vector of length 230, $A1$ and $A2$ are trainable elemental and space group attention weights, and $n$ and $q$ loop from 1 to $N_E$ and $N_{SG}$. The hidden material representation $h6_{m,n,q}$ is finally passed through two fully connected layers of $N_{h1}$ and $N_{h2}$ neurons to obtain the prediction for material property as:
\begin{eqnarray}
h7_{r} &=& f(b1_r + \sum_{m,n,q} W1_{m,n,q}^{r} \times h6_{m,n,q}) \label{eqn_full1}, \\
h8_{s} &=& f(b2_s + \sum_{r} W2_{r,s} \times h7_{r}) \label{eqn_full2}, \\
FE &=& b3 + \sum_{s} W3_s \times h8_s \label{eqn_full3},
\end{eqnarray}
where ($W1, b1$), ($W2, b2$), and ($W3, b3$) are trainable fully-connected layer weights and $r$ and $s$ vary from 1 to $N_{h1}$ and $N_{h2}$.

The neural network is implemented using openly available tensor library Tensorflow\cite{tensorflow}. The trainable weights are all initialized using a truncated Normal distribution of zero mean and unity standard deviation. The training is performed using the Adam optimizer\cite{kingma2014} with an initial learning rate of $1\times10^{-3}$. The training is performed for 10000 epochs with a batch size of 1024. The MAE is used as the loss function for the training of network. For training on top 20 most frequently appearing structure-types from OQMD, the loss on individual materials are scaled inversely by the number of materials in the structure type. The values of hyperparameters $N_{F1}$, $N_{F2}$ $N_W$, $N_E$, $N_{SG}$, $N_{h1}$, $N_{h2}$, and activation function are tested on the training error and the results are summarized in the Table~\ref{table_NN}.

The final network has a total of 11197 trainable weights which are obtained by setting the values of $N_{F1}$, $N_{F2}$, $N_W$, $N_E$, $N_{SG}$, $N_{h1}$, and $N_{h2}$ at 4, 4, 4, 8, 4, 60, and 10 respectively. The sigmoid function is used to introduce non-linearity in the network. The python code for generating U-apI descriptor from structure files and for performing ML using U-apI descriptor based ABCNN can be obtained through Github\cite{github}.

The prediction performance of U-apI descriptor based ABCNN is reported in Fig.~\ref{fig_top_prototype} for 20 most frequently appearing structure types from the OQMD. For these structure types, the VT apD descriptor based RF and the structure-type pre-clustering apI descriptor based RLFNN result in maximum MAEs of $0.10$ and $0.25$ eV/atom and a factor of five difference in the performance on sparsely and densely populated structure types. For the U-apI descriptor based ABCNN, with an exception of structure-type ($\text{ABC}_3$; 99; a,b,ac), the maximum MAE is $0.10$ eV/atom with only a factor of two difference in the performance across structure types.

As can be seen in Fig.~\ref{fig_validation}(b), for structure-type ($\text{ABC}_3$; 99; a,b,ac), several materials have FE as high as 5 eV/atom. These materials have compositions $\text{HfTlO}_3$, $\text{TmTlO}_3$, $\text{LuTlO}_3$, $\text{SmTlO}_3$, $\text{HoTlO}_3$, and $\text{OsTlO}_3$. All of these materials are erroneously reported to have identical lattice constants of $3.398$ and $3.639$ $\AA$ in the $a$ and $c$ directions in the relaxed configurations in the OQMD; thereby suggesting failed/ non-relaxed geometries. For these materials, the prediction errors are as high as 6 eV/atom. Removing these erroneous entries from the dataset brings down the MAE to below $0.10$ eV/atom for ($\text{ABC}_3$; 99; a,b,ac) structure-type using the U-apI descriptor based ABCNN (not shown in Fig.~\ref{fig_top_prototype}).

The ability of the U-apI descriptor based ABCNN to handle structurally diverse materials is tested by performing the learning on more than 428000 materials from OQMD. These materials belong to 90 different space groups with 289 participating Wyckoff-sites and 84 elements and are obtained by eliminating materials for which participating Wyckoff-sites appeared in fewer than 100 materials in the OQMD\cite{oqmd_screening}. On these materials, the U-apI descriptor based ABCNN results in a test MAE of only $0.07$ eV/atom.

The CNN has also been employed recently by Xie et al.~\cite{xie2018} with CG descriptor to predict the material properties of inorganic periodic solids.
When used in the prediction of FE of materials from the OQMD\cite{cgcnn_note}, this CG-based CNN results in a similar prediction performance as that of the VT apD descriptor based RF (Fig.~\ref{fig_top_prototype}). The CG based CNN, however, employs additional structural information about relaxed bond lengths of neighboring atoms which, similar to VT, requires relaxation of atomic coordinates and therefore, inhibits prediction for new materials. In contrast, the U-apI descriptor based ABCNN employs only space group and Wyckoff-sites of atoms and therefore  allows for screening of millions of compositionally different new materials within the same structure type.

\begin{figure}
\begin{center}
\includegraphics{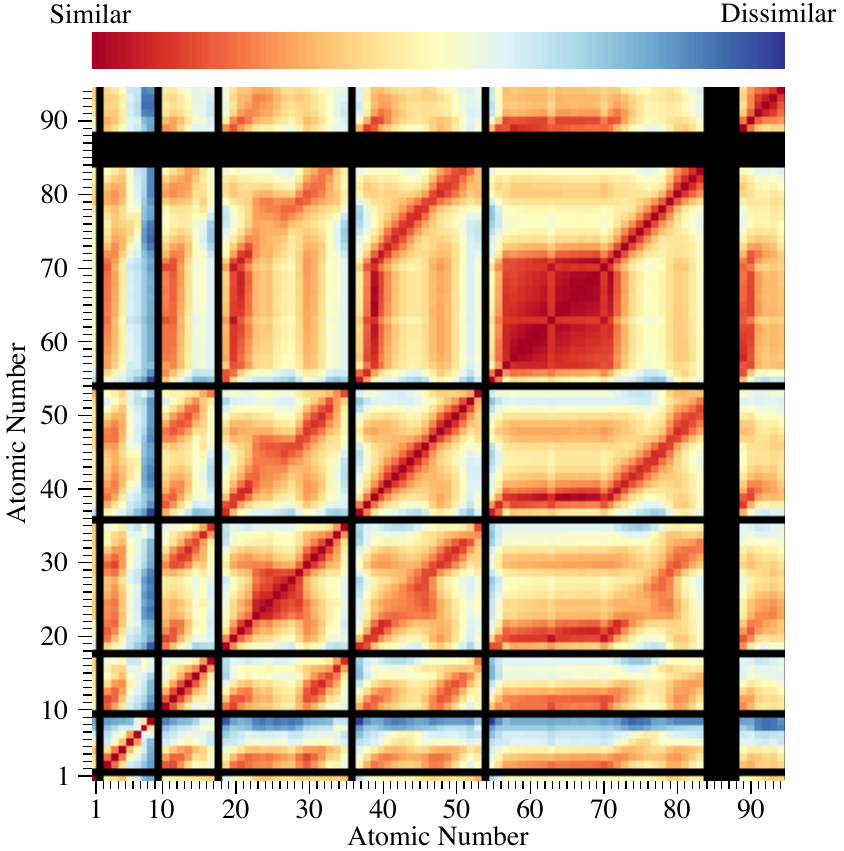}
\end{center}
\caption{{\bf Physical significance of attention weights.} The heatmap showing the similarity of elements to form chemical compounds. The similarities between elements are characterized using the Euclidean distance of elemental attention weights learned by the neural network. The non-participating elements are represented using black color.}
\label{fig_heatmap}
\end{figure}

The physical significance of attention-weights of elements in the U-apI descriptor based ABCNN is tested by identifying the similarities of elements to form chemical compounds. For this, the Euclidean distance is evaluated between elemental representation attention weights ($\text{A1}$ in Fig.~\ref{fig_NN}) as learned by the ABCNN. The resulting heatmap of distances is presented in Fig.~\ref{fig_heatmap}. As can be seen from Fig.~\ref{fig_heatmap}, the ABCNN learned correct elemental trends including the similarities of the same period elements of the periodic table to form the chemical compounds. Note that this heatmap is similar to that obtained by Glawe et al.\cite{glawe2016} using the statistical analysis of experimentally known materials; thereby establishing the physical relevance of attention-weights in the ABCNN.

In summary, we used crystallographic space group and Wyckoff-sites to describe structure details of materials in ML models. We find that space group and Wyckoff-sites fix the symmetry environment of materials where composition-only material descriptors are sufficient to distinguish different materials. We show that with sufficient materials in this fixed space group and Wyckoff-site space and with representation learning feedforward neural network, the minimal material descriptor employing only atomic numbers is able to match the performance of apD descriptor. Lastly, we present a U-apI descriptor which when used with attention-weights sharing convolution neural network, is able to learn across diverse  structure types and delivers a prediction MAE of $0.07$ eV/atom. This accurate and regularized performance of U-apI descriptor across diverse structure-types without explicit use of atomic positions (as in the case of other material descriptors), establishes the broader applicability of U-apI descriptor in the ML of material properties.

\begin{acknowledgments}
We thank Raul Abram Flores Jr. and Christopher Paolucci from Stanford University for their valuable feedback.
\end{acknowledgments}

\bibliography{references}

\begin{thebibliography}{27}
\bibitem [{cgc()}]{cgcnn_note}%
  \BibitemOpen
  \href@noop {} {}\bibinfo {note} {We used default (optimized) values of
  hyperparameters as provided by authors in their code.}\BibitemShut {Stop}%
\end{thebibliography}

\appendix*
\section{Hyperparameters Tuning}
\begin{table*}
  \caption{The effect of different hyperparameters on the training error of U-apI descriptor based ABCNN.}
  \label{table_NN}
  \begin{tabular}{c|c|c|c|c|c|c|c|c|c}
    \hline
    \multirow{2}{*}{$N_{F1}$} & \multirow{2}{*}{$N_{F2}$} & \multirow{2}{*}{$N_W$} & \multirow{2}{*}{$N_E$} & \multirow{2}{*}{$N_{SG}$} & \multirow{2}{*}{$N_{h1}$} & \multirow{2}{*}{$N_{h2}$} & Number of & \multirow{2}{*}{activation function} & Training Error\\
     & & & & & & & Trainable Parameters & & (eV/atom) \\
    \hline
    \hline
    4 & 4 & 4 & 4 & 4 & 20 & 10 & 3885 & Sigmoid & $0.10$ \\
    8 & 4 & 4 & 4 & 4 & 20 & 10 & 4425 & Sigmoid & $0.11$ \\
    4 & 8 & 4 & 4 & 4 & 20 & 10 & 4749 & Sigmoid & $0.12$ \\
    4 & 4 & 8 & 4 & 4 & 20 & 10 & 5597 & Sigmoid & $0.11$ \\
    4 & 4 & 4 & 8 & 4 & 20 & 10 & 5637 & Sigmoid & $0.08$ \\
    4 & 4 & 4 & 12 & 4 & 20 & 10 & 7389 & Sigmoid & $0.08$ \\
    4 & 4 & 4 & 4 & 8 & 20 & 10 & 6085 & Sigmoid & $0.11$ \\
    4 & 4 & 4 & 4 & 4 & 40 & 10 & 5385 & Sigmoid & $0.08$ \\
    4 & 4 & 4 & 4 & 4 & 60 & 10 & 6885 & Sigmoid & $0.09$ \\
    4 & 4 & 4 & 4 & 4 & 80 & 10 & 8385 & Sigmoid & $0.09$ \\
    4 & 4 & 4 & 4 & 4 & 20 & 20 & 4105 & Sigmoid & $0.11$ \\
    4 & 4 & 4 & 4 & 4 & 20 & 10 & 3885 & Elu & $0.11$ \\
    4 & 4 & 4 & 4 & 4 & 20 & 10 & 3885 & Softplus & $0.12$ \\
    4 & 4 & 4 & 4 & 4 & 20 & 10 & 3885 & Softsign & $0.10$ \\
    4 & 4 & 4 & 4 & 4 & 20 & 10 & 3885 & Tanh & $0.10$ \\
    \hline
  \end{tabular}
\end{table*}

\end{document}